\documentclass[aip,apl,reprint,floatfix,superscriptaddress,amsmath,showpacs]{revtex4-1}

\usepackage{graphicx}
\usepackage{dcolumn}
\usepackage{bm}
\usepackage{amssymb}
\usepackage{wrapfig}
\usepackage{textcomp} 

\begin{document}

\title{Microwave-induced spin currents in ferromagnetic-insulator$|$normal-metal bilayer system}

\author{Milan Agrawal}
\email{magrawal@physik.uni-kl.de}
\affiliation{Fachbereich Physik and Landesforschungszentrum OPTIMAS, Technische Universit\"at Kaiserslautern, 67663 Kaiserslautern, Germany}
\affiliation{Graduate School Materials Science in Mainz, Gottlieb-Daimler-Strasse 47, 67663 Kaiserslautern, Germany }

\author{Alexander  A. Serga}
\affiliation{Fachbereich Physik and Landesforschungszentrum OPTIMAS, Technische Universit\"at Kaiserslautern, 67663 Kaiserslautern, Germany}

\author{Viktor Lauer }
\affiliation{Fachbereich Physik and Landesforschungszentrum OPTIMAS, Technische Universit\"at Kaiserslautern, 67663 Kaiserslautern, Germany}

\author{Evangelos Th. Papaioannou}
\affiliation{Fachbereich Physik and Landesforschungszentrum OPTIMAS, Technische Universit\"at Kaiserslautern, 67663 Kaiserslautern, Germany}

\author{Burkard Hillebrands}
\affiliation{Fachbereich Physik and Landesforschungszentrum OPTIMAS, Technische Universit\"at Kaiserslautern, 67663 Kaiserslautern, Germany}

\author{Vitaliy I. Vasyuchka}
\affiliation{Fachbereich Physik and Landesforschungszentrum OPTIMAS, Technische Universit\"at Kaiserslautern, 67663 Kaiserslautern, Germany}

\date{\today}

\begin{abstract}

A microwave technique is employed to simultaneously examine the spin pumping and the spin Seebeck effect processes   in a YIG$|$Pt bilayer system. The experimental results show that for these two processes, the spin current  flows in opposite directions. The temporal dynamics of the longitudinal spin Seebeck effect exhibits that the effect depends on the diffusion of bulk thermal-magnons in the thermal gradient in the ferromagnetic-insulator$|$normal-metal system.

\end{abstract}


\maketitle

Since its discovery in 2008, the spin Seebeck effect \cite{Uchida2008}---a route to generate a spin current by applying a heat current to ferromagnets---has given a new dimension to the field of spin-caloritronics\cite{Bauer2012}.  In particular, the longitudinal spin Seebeck effect (LSSE) \cite{Uchida2010b}, where the spin current flows along the thermal gradient in the magnetic material, drives the field due to its technologically  promising applications in energy harvesting \cite{Kirihara2012}, and in temperature, temperature gradient, and position sensing\cite{Uchida2011a}. 

Having conceptual understanding and future applications   in the centre of attention, a comparative  study of the spin current direction and its temporal evolution for different spin-current-generation processes like spin pumping (SP) and spin Seebeck effect (SSE) is very important. In previous experiments \cite{Jungfleisch2013, Sandweg2011}, these issues have not been explicitly  addressed. In this letter, we demonstrate microwaves as a simple-and-controlled tool to investigate both, SP and SSE, processes simultaneously in a single experiment. Such investigations are not possible with other techniques including laser heating \cite{Agrawal2014a}  or direct-current heating \cite{Schreier2013a} employed to study the SSE. For example, in  Ref.~\onlinecite{Schreier2014}, the direction of the spin current in the SP and SSE processes has been determined by combining the FMR technique with additional Peltier or dc/ac based heating techniques. Our results reveal that in a ferromagnet$|$normal metal (paramagnet) system , the spin current flows from the ferromagnet (FM) to the normal metal (NM) in the case of SP process, while the flow reverses for the LSSE provided  that the NM is hotter than the FM. The time-resolved measurements show  that the spin current dynamics of the LSSE is on sub-microsecond timescale compared to nanosecond fast spin pumping process \cite{Jungfleisch2011}.

The experiment was realized using a bilayer of a magnetic insulator, Yttrium Iron Garnet (YIG), and a normal metal, Pt. The sample structure consists a 6.7-$\mu\rm m$-thick YIG film of dimensions $\rm{14~mm \times 3~mm}$, grown by liquid phase epitaxy on a 500-$\mu \rm m$-thick Gallium Gadolinium  Garnet (GGG)  substrate, and a 10-$\rm{nm }$-thick Pt strip		 ($\rm{3~mm \times 100~ \mu m}$), structured by photolithography and deposited   by molecular beam epitaxy at a growth rate of $0.05$~\AA$\rm /s$ under a pressure of $5 \times10^{-11} ~\rm {mbar}$. A  0.6-mm-wide and  17-$\mu \rm m$-thick copper (Cu) microstrip antenna was designed on a dieletric substrate to apply microwaves to the sample.  In order to obtain  a maximum microwave heating efficiency by eddy currents in the metal (Pt), the Pt-covered surface of the sample was placed on top of the micro-stripline. An insulating layer was inserted in between the sample and the micro-stripline to avoid any direct electric contact. Gold wires were glued to the Pt strip with silver paste  to connect with the external circuit. 
%Figure 1
\begin{figure}[t]
	  \begin{center}
    \scalebox{1}{\includegraphics[width=\columnwidth, clip]{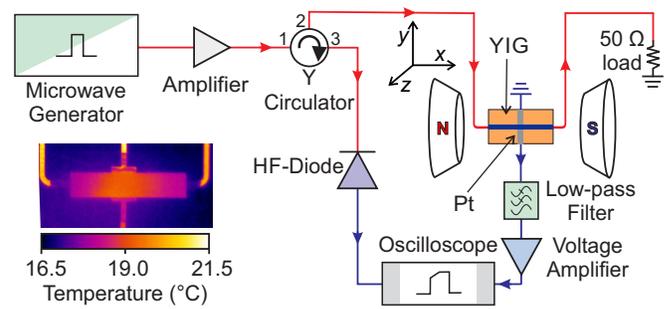}}
    \end{center}
	  \caption{\label{setup} A schematic sketch of the experimental setup. Microwaves were employed to heat a 10-nm-thick Pt strip grown on a 6.7-$\mu$m-thick YIG film placed on top of the micro-stripline. The reflected microwaves were monitored on an oscilloscope. The inverse spin Hall  voltage generated in the Pt strip was amplified   and measured by the oscilloscope. The inset is an infrared thermal image of the sample obtained by continuous microwave heating of the Pt strip situated in the middle of the film.}
\end{figure}

A schematic diagram of the experimental setup is shown in Fig.~\ref{setup}.  Microwaves from an Anritsu MG3692~C generator were amplified (+30~dB) by an amplifier and guided to the sample structure. The microwaves are absorbed by the thin Pt metal strip and start to heat it up. As a result, a thermal gradient along the $+\,z$-direction was established.  The reflected microwaves from the micro-stripline were received by connecting a Y-circulator to the microwave circuit-line. The reflected microwaves were rectified using a high-frequency (HF) diode and monitored on an oscilloscope. This signal provides the information about the shape of  the microwave signal envelope and its duration. The YIG film was magnetized in plane by an applied magnetic field $\mu_0H$ along the $x$-axis.

The perpendicular thermal gradient in the YIG$|$Pt bilayer generates a spin current in the system due to the LSSE. The generated spin current flows along the thermal gradient ($z$-axis). In the normal metal Pt, the spin current converts into a charge current  by the inverse spin Hall effect (ISHE) \cite{Saitoh2006} as $\textbf{J}_\mathrm{ISHE}  \propto {\textbf{J}}_\mathrm{s} \times {\boldsymbol \sigma}, $ where $\textbf{J}_\mathrm{s}$  is the spin current, and $\boldsymbol{\sigma}$ the spin polarization. The generated charge current along the $y$-axis in the Pt strip  was passed through a  low-pass filter to block  alternating currents generated directly by the electric field components of the microwaves. The filtered signal was amplified and observed on the oscilloscope.

\begin{figure}[t]
	  \begin{center}
    \scalebox{1}{\includegraphics[width=\columnwidth, clip]{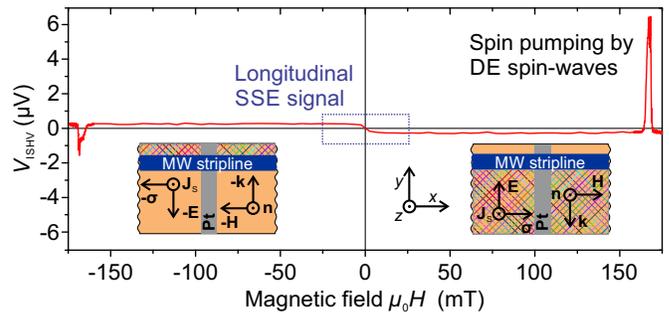}}
    \end{center}
	  \caption{\label{figure2} The inverse spin Hall voltage ($V_\mathrm{ISHE}$) generated in the Pt strip as a function of the applied magnetic field $\mu_0 H$. An asymmetry in the amplitude of $V_\mathrm{ISHE}$ at FMR-magnetic field appears due to unequal efficiency of Damon-Eshbach spin-waves excitation  (shaded area) for two opposite directions of the applied magnet field, shown in the insets.}
\end{figure}

In the first experiment, continuous microwave measurements at a fixed frequency of $\rm 6.8~GHz$ were carried out  by varying the  magnetic field. In Fig.~\ref{figure2}, the inverse spin Hall voltage $V_{\rm ISHE}$ is plotted versus the applied magnetic field $\mu_0 {H}$. 
Clearly, three features can be noticed here: (i) two peaks with opposite polarities at the magnetic fields of $+\,168.6$~mT and $-\,168.6$~mT, (ii) their unequal magnitude, and (iii) an offset for all non-resonance magnetic fields, which has an opposite polarity to the peaks. The first feature originates from the spin pumping process by spin waves excited close to ferromagnetic resonance (FMR) in the YIG film \cite{Kajiwara2010, Costache2006, Ando2009b}. The FMR conditions were achieved for both positive and negative magnetic fields. Corresponding to these fields, a spin current is injected into Pt by the spin pumping process. The inverse spin Hall voltage generated in Pt is given by  $V_\mathrm{ISHE}  \propto \theta_\mathrm{SHE} \,({\textbf{J}}_\mathrm{s} \times {\boldsymbol \sigma}) $, where  $\theta_\mathrm{SHE}$ denotes the spin Hall angle. The direction of $\boldsymbol{\sigma}$ depends on the direction of the magnetic field. Therefore, on inverting the magnetic field direction, the polarity of $ {V}_\mathrm{ISHE}$ reverses.

%Figure 2

In order to understand the second feature of the spectrum that the signals have unequal amplitude, it is  important to discuss the spin-wave modes excited in the YIG film for our experimental geometry. It is clear from the sample orientation, shown in Fig.~\ref{setup} and inset to Fig.~\ref{figure2}, that the spin waves excited  by the Oersted field of the micro-stripline propagate along the $y$-axis, perpendicular to the  magnetic field applied along the $x$-axis. These kinds of spin waves with wave vector $\textbf{k}\bot \textbf{H}$ are known as magnetostatic surface spin waves (MSSW) or Damon-Eshbach (DE) spin-waves \cite{Damon1961}. The DE spin-waves are nonreciprocal spin waves and travel along a direction given by $ \textbf{k}=\textbf{H}\times\textbf{n} $, where $\textbf{n}$ is the normal to the film surface. Therefore, the propagation of these spin waves on the surface of a film can be reversed  by inverting the direction of the magnetic field.

In our  experiment, when a magnetic field is applied along the $+\,x$-direction, The DE-spin waves, in the YIG film surface close to the micro-stripline \cite{Note1}, can only be excited  along the $-\,y$-direction ($\rm \hat{\textbf{k}}={\textbf{x}} \times {\textbf{z}}$) with respect to the micro-stripline as shown in the inset to Fig~\ref{figure2}. On the other hand, when the magnetic field is applied along the  $-\,x$-direction, the spin waves can propagate only along the  $+\,y$-axis. If the YIG film is not positioned symmetrically around the micro-stripline, as in our case, the effective YIG film area, where spin waves can be excited,  will be unequal for two opposite fields as shown in the inset to Fig~\ref{figure2}. Since the strength of  $V_{\rm ISHE}$ signal is proportional to the spin-wave intensity in the system \cite{Sandweg2011}, we observed an unequal amplitude of $V_{\rm ISHE}$ in our experiment. We performed alike measurements with displacing the YIG film and find that the amplitude of the spin pumping signals can be altered by varying the relative positions of the film with respect to the micro-stripline.  Therefore, we conclude that the unidirectional nature of the DE spin-waves regulates the asymmetry of the ISHE signal \cite{Iguchi2013}.

The third feature seen in Fig.~\ref{figure2}, i.e., an offset for non-resonant  magnetic fields; is attributed to the LSSE. A similar signal could also be produced by the anomalous Nernst effect in Pt, magnetized due to the proximity effect. However, recent observations \cite{Geprags2012, Kikkawa2013} discard any such possibility in YIG$|$Pt systems.   The polarity of the LSSE signal changes with the direction of the magnetic field; however, it is important to notice that  the LSSE signal has an opposite polarity than that of the signal at FMR for a same direction of the magnetic field. This evidence excludes the possibility of  non-resonant spin pumping in the system. When the Pt strip is heated by microwave absorption, a thermal gradient ($\nabla T_{\rm z}$) from YIG to Pt develops normal to the interface. The thermal gradient generates a spin current flowing along the $z$-axis. Since the Pt strip is hot, the spin current generated via the longitudinal SSE ($\textbf{J}_{\rm s} \propto - \boldsymbol \nabla \rm T$) flows from Pt to YIG \cite{ Xiao2010,Rezende2014}, in contrast to spin pumping where the spin current flows from YIG to Pt \cite{Tserkovnyak2002a}. This argument explains the opposite polarities of the resonant (spin pumping) and the non-resonant (longitudinal SSE) inverse-spin-Hall voltages ($V_\mathrm{ISHE}$) observed in Fig.~\ref{figure2}. These results are consistent with previous experimental studies \cite{Jungfleisch2013, Sandweg2011,Schreier2014}. As discussed above, the non-resonant $V_\mathrm{ISHE}$ is attributed to the longitudinal SSE; henceforth, we denote the non-resonant $V_\mathrm{ISHE}$ values as  $V_\mathrm{LSSE}$. 

%Figure 3
\begin{figure}
	  \begin{center}
    \scalebox{1}{\includegraphics[width=\columnwidth, clip]{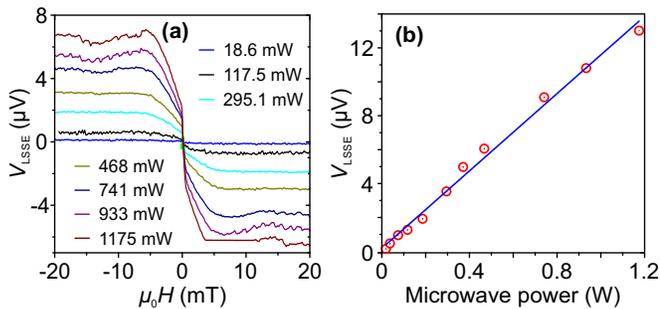}}
    \end{center}
	  \caption{\label{figure3} (a) Plotted is $V_\mathrm{LSSE}$ as a function of the  applied magnetic field for various applied microwave powers. (b) The peak-to-peak amplitude of $V_\mathrm{LSSE}$ is plotted versus the applied microwave power. The  peak-to-peak amplitude of $V_\mathrm{LSSE}$ scales linearly with the microwave power.}
\end{figure}

Magnetic field scans for various microwave input powers were carried out. In Fig.~\ref{figure3}, the $V_\mathrm{LSSE}$ versus the magnetic field data is plotted for various applied microwave powers.  With increasing microwave power, the temperature increases in the Pt strip which enlarges the thermal gradient close to the YIG$\vert$Pt  interface and, hence, injects a larger spin current ($\textbf{J}_{\rm s} \propto - \boldsymbol \nabla \rm T$) into the YIG film \cite{Uchida2010b, Xiao2010, Rezende2014}. Impact of the large spin current appears as a higher $V_\mathrm{LSSE}$ signal highlighted in Fig.~\ref{figure3}(b). The peak-to-peak amplitude of $V_\mathrm{LSSE}$ scales linearly with the applied microwave power. The signature that the $V_\mathrm{LSSE}$ signal scales linearly with the microwave power verifies that the signal originates from the heating produced in Pt shown in the inset to Fig.~\ref{setup}.

The above  experiment demonstrates that microwaves can be utilized to create a thermal gradient in ferromagnetic-insulator$|$normal-metal system, thereby, to study the LSSE along with the SP. The experimental setup shown in Fig.~\ref{setup} can also be employed to investigate the temporal dynamics of the longitudinal SSE and to compare it with the SP dynamics \cite{Jungfleisch2011}. In the second experiment, instead of continuous microwaves, 10-$\mu s$-long microwave pulses with rise-fall times of less than 10~ns were used to perform the  time-resolved measurements of the longitudinal SSE. %of power $\rm \approx 0.32\,mW$ ($\rm 25\,dBm$). 
The frequency of the microwave pulses ($\rm 6.8~GHz$) was chosen such that the magnetic system stayed at non-resonance condition  of the magnetic field  in the range of interest ($\rm \pm\,25~mT$). The experiment was executed at various microwave powers. %to get a better LSSE signal 
The measurements were recorded for both positive (+\,25~mT) and negative (-\,25~mT) magnetic fields, and an average value of the non-resonant $V_\mathrm{ISHE}$, i.e., $V_\mathrm{LSSE}$  was considered. 
 
%Figure 4
\begin{figure}[b]
	  \begin{center}
    \scalebox{1}{\includegraphics[width=\columnwidth, clip]{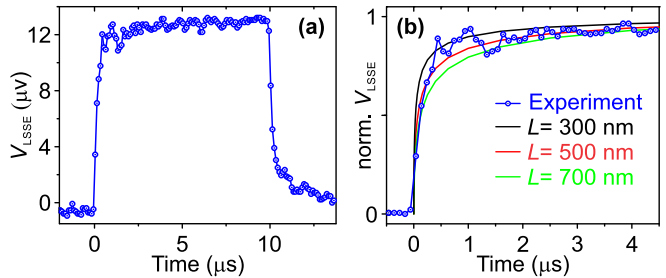}}
    \end{center}
	  \caption{\label{figure4} (a) Plotted is the temporal evolution of $V_\mathrm{LSSE}$ on the application of  a $10~ \mu \rm s$  long microwave pulse which creates a vertical temperature gradient in the YIG$\vert$Pt structure by heating the Pt strip. (b) A comparison of experimentally measured $V_\mathrm{LSSE}$ data with calculated values using Eq.~(\ref{eq1}) for various effective magnon diffusion lengths $L=300,~500,~700~ \rm{nm}$.}
\end{figure}

In Fig.~\ref{figure4},   $V_\mathrm{LSSE}$ is plotted versus  time.  The longitudinal SSE signal takes around $1~ \mu \rm s $ to reach to the saturation level. The $10\%-90\%$ rise time of the signal is found to be around $530~\rm ns$. The longitudinal SSE signal ($V_\mathrm{LSSE}$) shows similar features as reported in Ref.~\onlinecite{Agrawal2014a}, where a pulsed laser is employed to create the vertical thermal gradient in the YIG$|$Pt system. The main  difference observed here is that  the $V_\mathrm{LSSE}$ signal appears as soon as the microwave current runs; contrarily,  in the laser heating experiment \cite{Agrawal2014a} a time lag (200~ns) exists due to the laser switching time.

The model of  thermal magnon diffusion\cite{Agrawal2014a, Rezende2014, Ritzmann2014} is employed here to understand the timescale of the longitudinal SSE. The model states that the spin current from a FM  injected  into a NM depends on the diffusion of thermal magnons in the FM. The density of thermal magnons is proportional to the local phonon temperature \cite{Agrawal2013,Schreier2013}. Due to the thermal gradient in the FM, magnons diffuse from hotter regions (higher population) to colder regions (lower population) of the FM and create a magnon density inequilibrium at the FM$|$NM interface which leads to the injection of a spin current into the NM \cite{Xiao2010}. The timescale of the effect depends on the temporal development of the magnon density inequilibrium, i.e., the  thermal gradient in the system. According to the model, $V_\mathrm{LSSE}$ is given by\cite{Agrawal2014a}
\begin{equation}
\label{eq1}
V_\mathrm{LSSE}(t)\propto   \int \limits_{\rm interface}^{l} \nabla T_z(z,t) \exp(\frac{- \left|z\right|}{L}) dz,
\end{equation}
where  $\nabla T_z$  is the phonon thermal gradient  in the FM, perpendicular to the interface, \textit{l} is the magnetic film thickness, and \textit{L} is the effective magnon diffusion length. 

We fitted our experimental data shown in Fig~\ref{figure4}(a) with Eq.~\ref{eq1} using $\nabla T_z(z,t)$ calculated numerically  by solving the heat equation for our system in accordance with a model described in Ref.~\onlinecite{Agrawal2014a}. In Fig.~\ref{figure4}(b), the normalized experimental $V_\mathrm{LSSE}$-signal was plotted together with the calculated ones for various magnon diffusion lengths of $300~ \rm nm$, $500 ~ \rm nm$, and $700~ \rm nm$. The model resembles the experimental data well. The fitting shows that a typical magnon diffusion length for thermal magnons in the YIG$|$Pt system is around $500 ~ \rm nm$. An identical value for the magnon diffusion length was obtained in the laser heating experimental performed on the same sample, reported in Ref.~\onlinecite{Agrawal2014a}.

In summary, we presented microwaves as a perspective heating technique to generate a thermal gradient in  ferromagnetic insulator$|$normal metal systems to study the static and temporal dynamics of the longitudinal spin Seebeck effect. The static measurements provide  crucial information about the direction of the spin current flow in the spin pumping and longitudinal SSE processes. The experiment demonstrates that in the longitudinal SSE a spin current flows from the normal metal (hot) towards the ferromagnet (cold) while in the spin pumping case, the flow  is opposite. The temporal dynamics of the longitudinal SSE experiment manifests the sub-microsecond timescale of the effect which is slower than the spin pumping process. The thermal magnon diffusion model can explain the outcomes of the experiment and leads to conclude that the timescale of the effect relies the evolution of the vertical thermal gradient in the vicinity of the ferromagnet$|$normal metal interface. From our experiment, a typical magnon diffusion length of $ 500 ~ \rm nm$ is estimated  for the YIG$|$Pt system.

The authors thank A. V. Chumak, M. B. Jungfleisch, and P. Pirro for valuable discussions. M.A. was supported by a fellowship of the Graduate School Material Sciences in Mainz (MAINZ) through DFG funding of the Excellence Initiative (GSC-266). We acknowledge financial support by Deutsche Forschungsgemeinschaft (SE 1771/4) within Priority Program 1538 ``Spin Caloric Transport", and  technical support from the Nano Structuring Center, TU~Kaiserslautern.

\end{document}